\documentclass[12pt,preprint]{aastex}
\shorttitle{NIR Extinction due to Cool SN Dust in Cas A}
\shortauthors{Lee et al.}

\usepackage{amsmath}
\usepackage{amsfonts}
\usepackage{amssymb}
\usepackage{color}
\usepackage{multirow}

\newcommand{\msun}{M_{\sun}}
\newcommand{\kms}{{\rm km~s}^{-1}}
\newcommand{\hei}{He {\small I}}

\newcommand{\oi}{[\ion{O}{1}]}
\newcommand{\oiii}{[\ion{O}{3}]}
\newcommand{\oiv}{[\ion{O}{4}]}
\newcommand{\neii}{[\ion{Ne}{2}]}
\newcommand{\neiii}{[\ion{Ne}{3}]}
\newcommand{\arii}{[\ion{Ar}{2}]}

\newcommand{\silii}{[\ion{Si}{2}]}

\newcommand{\sii}{[\ion{S}{2}]}
\newcommand{\siii}{[\ion{S}{3}]}
\newcommand{\siv}{[\ion{S}{4}]}
\newcommand{\feii}{[\ion{Fe}{2}]}

\begin{document}

\title{Near-infrared Extinction due to Cool Supernova Dust in Cassiopeia A}

\author{
Yong-Hyun Lee\altaffilmark{1},
Bon-Chul Koo\altaffilmark{1},
Dae-Sik Moon\altaffilmark{2},
and Jae-Joon Lee\altaffilmark{3}}

\email{yhlee@astro.snu.ac.kr}
\altaffiltext{1}{Department of Physics and Astronomy, Seoul National University,
Seoul 151-742, Korea}
\altaffiltext{2}{Department of Astronomy and Astrophysics, University of Toronto,
Toronto ON M5S 3H4, Canada}
\altaffiltext{3}{Korea Astronomy and Space Science Institute,
Daejeon 305-348, Korea}

%%%%%%%%%%%%%%%%%%%%%%%%%%%%%%%%%%%%%%%%%%%%%%%%%%%%%%%%%%%%%%%%%%%%%%%%%%%80==>
\begin{abstract}

We present the results of extinction measurements
toward the main ejecta shell of the Cassiopeia A supernova (SN) remnant
using the flux ratios between the two near-infrared (NIR) 
\feii\ lines at 1.26 and 1.64 \micron.
We find a clear correlation
between the NIR extinction ($E(J-H)$) and the radial velocity of ejecta knots,
showing that redshifted knots are systematically more obscured 
than blueshifted ones.
This internal ``self-extinction'' strongly indicates that a large amount of 
SN dust resides inside and around the main ejecta shell.
At one location in the southern part of the shell, 
we measure $E(J-H)$ by the SN dust of $0.23\pm0.05$ mag.
By analyzing the spectral energy distribution of 
thermal dust emission at that location, we show that
there are warm ($\sim 100$~K) and cool ($\sim 40$~K) SN dust components 
and that the latter is responsible for the observed $E(J-H)$.
We investigate the possible grain species and size of each component 
and find that the warm SN dust needs to be silicate grains
such as MgSiO$_{3}$, Mg$_{2}$SiO$_{4}$, and SiO$_{2}$,
whereas the cool dust could be either 
small ($\lesssim 0.01~\micron$) Fe or large ($\gtrsim 0.1~\micron$) Si grains.
We suggest that 
the warm and cool dust components in Cassiopeia A represent 
grain species produced in diffuse SN ejecta and in dense ejecta clumps,
respectively.

\end{abstract}


%%%%%%%%%%%%%%%%%%%%%%%%%%%%%%%%%%%%%%%%%%%%%%%%%%%%%%%%%%%%%%%%%%%%%%%%%%%80==>
\section{Introduction} \label{sec-int}

For the past decade,
evidence for copious amounts of dust ($\ge10^{8}~\msun$)
in high-redshift galaxies and quasars with $z>5$
has been reported in many far-infrared (FIR) and
sub-millimeter/millimeter studies
\citep[e.g.,][and references therein]{wan11,lei13,cal14},
implying that dust should be produced within a very short timescale,
less than 1 Gyr after the Big Bang.
The most promising sites of dust formation in such an early universe 
are believed to be core-collapse supernovae (CCSNe),
since their progenitors have very short lifetimes ($\sim10^6$ years) 
compared to asymptotic giant branch stars,
which are major sources of dust in the local universe \citep{mor03,mar06}.

To understand the dust formation process in expanding supernova (SN) ejecta,
many theoretical studies have been conducted to date
\citep{noz03,noz08,noz10,sar14}.
According to these theoretical studies, dust can form 
in an expanding He core with grain species determined by 
the elemental composition at the formation site. 
If an SN ejecta maintains its layered structure during expansion,
a variety of grain species can form,
e.g., carbon grains in the helium layer,
silicate and oxide grains in the oxygen-rich layer, and
Si/Fe grains in the innermost layer,
the amounts of which depend on the progenitor mass.
If the SN ejecta is mixed, however,
it is mostly silicate and oxide grains that can form.
The size of the grains depends on the gas density at the condensation time.
In an SN IIP, which has a thick hydrogen envelope, the expansion is slow,
so large grains with an average radius of $\sim 0.1~\micron$
can form \citep{noz03}.
In an SN IIb with a low-mass hydrogen envelope as well as 
envelope-stripped SNe such as Type Ib/c, on the other hand, 
the SN ejecta expands rapidly,
so only small grains with a size less than 0.01 \micron\
can form \citep{noz08,noz10}.
The size distribution and chemical composition of the dust, however,
could be greatly affected by density enhancements
due to clumping \citep{sar14}.
Theoretical studies predict that
one CCSN can produce 0.1--$1.0~\msun$ of SN dust 
\citep{koz91,tod01,noz03,noz08,noz10,bia07}, but 
the dust masses obtained from observations of nearby CCSNe
are much smaller than this, e.g., $\lesssim 10^{-2}~\msun$
\citep{mei07,kot09,sak09,sza11}.
A significant ($\gtrsim 0.1~\msun$) amount of dust has been observed in
several young CC supernova remnants (SNRs), e.g., 
Cassiopeia A \citep{bar10,sib10,are14}, Crab \citep{gom12,owe15},
and 1987A \citep{mat11,ind14},
but the physical and chemical properties of the dust remain to be explored
(see below).

Cassiopeia A (Cas A),
which is a remnant of an SN IIb explosion \citep{kra08},
is an ideal target for examining the properties of SN dust.
It is quite young \citep[$\sim 330$ years;][]{fes06},
so the SN ejecta and freshly formed SN dust have not been 
significantly mixed with the circumstellar/interstellar medium (CSM/ISM),
and it is relatively close \citep[3.4 kpc;][]{ree95},
so the physical and chemical structures can be resolved. 
Therefore, Cas A has been a major target of infrared space missions.
The first direct evidence for SN dust in Cas A
came from the {\it Infrared Space Observatory} ({\it ISO}),
which performed mid-infrared (MIR; 2.4--45 \micron)
spectroscopic observations toward the bright ejecta shell
\citep{lag96,are99,dou01}.
The continuum spectra had a strong bump peaking at 21 \micron\
together with a relatively weak bump at 9.5 \micron, and 
\citet{dou01} showed that the spectra can be fitted by 
two dust components at different temperatures,
one at 90 K and the other at 350 K,  
with pyroxene (MgSiO$_3$), quartz (SiO$_2$), 
and aluminum oxide (Al$_2$O$_3$) as major components. 
The composition and distribution of this ``warm'' SN dust
heated by a reverse shock have been studied in detail using the 
spectral mapping data of the {\it Spitzer Space Telescope} 
\citep{enn06,rho08,are14}.
These studies showed that the dust emission exhibits
distinct spectral characteristics depending on the 
SN material with which it is associated.
The most prominent dust emission 
is that with strong 9 and 21 \micron\ bumps,
which is associated with SN ejecta having strong Ar emission lines.
The spectral features agree with the {\it ISO} spectra and 
can be reproduced by Mg protosilicate/MgSiO$_3$,  
Mg$_{0.7}$SiO$_{2.7}$, or SiO$_2$. 
The dust emission associated with SN ejecta 
having strong Ne lines is smooth without any silicate features 
and can be reproduced by Al$_2$O$_3$ dust or C glass. 
There is yet another type of smooth dust emission
associated with X-ray Fe emission, but 
it is probably from swept-up CSM \citep{are14}.
Meanwhile, \citet{noz10} showed that
the observed spectral energy distribution (SED) of dust emission
can be fitted by a physical model
in which the SNR is expanding into a dense CSM,
and the MIR emission is mostly from silicate grains
(e.g., MgSiO$_{3}$, Mg$_{2}$SiO$_{4}$, and SiO$_2$) and MgO. 
The estimated total mass of warm dust ranges from 0.008 to 0.054 $\msun$. 

A larger amount of SN dust, however, appears to reside in 
the interior of Cas A, where the SN ejecta is freely expanding.
The first report was made by 
\citet{dun03}, who claimed the detection of 2--4 $\msun$ of
cold (15--20 K) SN dust associated with Cas A from 
an 850 \micron\ observation.
It was found, however, that at least some of the 850 \micron\ emission is 
from a foreground material \citep{kra04, dun09}, 
and instead \citet{sib10} and \citet{bar10}, 
using the balloon-borne sub-millimeter telescope BLAST and
infrared space telescopes {\it AKARI} and {\it Herschel},
showed that there is faint FIR and sub-millimeter emission from Cas A 
that can be fitted by dust at 35 K with a total mass of $\sim 0.07~\msun$.
\citet{are14} also identified the FIR emission from this ``cool'' dust 
associated with unshocked SN ejecta (their ``\silii'' dust group)
using the {\it Herschel} FIR data
and derived an upper limit of 0.1 $\msun$ for the dust mass.
However, the composition of the cool dust has been undetermined so far 
because the absorption cross sections of dust grains
in the FIR are mostly smooth without prominent spectral features.

In this paper, we show that one can explore the characteristics of 
the cool SN dust in the interior of Cas A using extinction in 
the near-infrared (NIR). 
Recently, we performed NIR spectroscopy toward the main ejecta shell of 
Cas A, where we obtained the spectral and kinematical properties of 
63 knots bright in \feii\ emission lines
\citep{koo13}.
%(@Lee et al. 2015; hereafter Paper I, see also @Koo et al. 2013).
Most of these knots are SN material, and the extinctions toward these knots 
show clear evidence for the SN dust, which can be 
combined with thermal infrared dust emission to infer the dust composition.
This paper is organized as follows.
In Section~\ref{sec-obs},
we outline our NIR spectroscopic observations and data reduction
and then briefly describe the public MIR and FIR imaging data
that we used in the paper.
In Section~\ref{sec-ext}, we derive the NIR extinction toward the knots
using \feii\ line ratios and show that
there is a correlation between the extinction and the line-of-sight velocity,
which implies extinction by SN dust.
Then, in Section~\ref{sec-sed},
by comparing the NIR extinction and the thermal dust infrared SED,
we constrain the possible compositions and sizes of the SN dust.
We discuss the implication of our result for dust production
in Cas A in Section~\ref{sec-dis} and
summarize our paper in Section~\ref{sec-sum}.

%%%%%%%%%%%%%%%%%%%%%%%%%%%%%%%%%%%%%%%%%%%%%%%%%%%%%%%%%%%%%%%%%%%%%%%%%%%80==>
\section{Observations and Data Reduction} \label{sec-obs}
%%%%%%%%%%%%%%%%%%%%%%%%%%%%%%%%%%%%%%%%%%%%%%%%%%%%%%%%%%%%%%%%%%%%%%%%%%%80==>
\subsection{Near-infrared Spectroscopy} \label{sec-obs-nir}

Eight long-slit NIR spectra of the main SN ejecta shell
were obtained on 2008 June 29 and August 8
using TripleSpec mounted on the Palomar Hale 5 m telescope;
this spectrograph provides a broadband spectrum
simultaneously covering 0.94--2.46 \micron\
with a moderate resolving power of $\sim 2700$.
We also observed the spectra of an A0V standard star (HD 223386)
for flux calibration just before and after the target observations.
To subtract sky OH airglow emission as well as background continuum,
the observations were made with either a consecutive ABBA pattern or
On-Off mode depending on the complexity of the field.
Figure~\ref{fig-slit}(a) shows the position of the slits
on the continuum-subtracted \feii\ 1.64 \micron\ narrow-band image
obtained in 2008 August.

We followed a general reduction procedure
to make a two-dimensional dispersed image.
Numerous sky OH airglow emission lines
were used for wavelength calibration,
and then we corrected it to the heliocentric reference frame.
The absolute photometric calibration was made by
comparing the observed spectrum of the standard star with
the Kurucz model spectrum \citep{kur03}.
Even though the absolute flux uncertainty goes up to 30\% depending on
the centering accuracy of the standard star observation,
the relative fluxes are quite robust.

We found a total of 63 infrared knots
using a clump-finding algorithm \citep{wil94}
and performed a single Gaussian fit for all the detected emission lines
to derive their fluxes and line-of-sight velocities.
For detailed descriptions of the data reduction and analysis
as well as the full list of parameters of the detected emission lines,
please refer to \citet{koo13} and Y.-H. Lee et al. (2015, in preparation).

%%%%%%%%%%%%%%%%%%%%%%%%%%%%%%%%%%%%%%%%%%%%%%%%%%%%%%%%%%%%%%%%%%%%%%%%%%%80==>
\subsection{Mid- and Far-infrared Imaging Data} \label{sec-obs-mir}

For an SED analysis of SN dust, 
we fully exploit the previously published archival data
providing the highest spatial resolution and sensitivity
in the MIR and FIR wavebands covering 10--500 \micron,
where the dust emission is dominant.

For short wavebands,
we use the W3 channel image ($\lambda_{\rm iso} \sim 12~\micron$) of
the {\it Wide-field Infrared Survey Explorer} \citep[{\it WISE}:][]{wri10}
with an angular resolution of $6\farcs5$.
We first retrieved all 36 single-exposure images (``Level-1b'') from the online
service\footnote{\url{http://irsa.ipac.caltech.edu/applications/wise/}}
and converted the pixel unit to MJy sr$^{-1}$.
After subtracting the predicted brightness of zodiacal light
calculated from the interplanetary dust model from each frame \citep{kel98},
we co-added the individual frames
using Montage software\footnote{\url{http://montage.ipac.caltech.edu/}}
to obtain the final image.
Instead of the {\it WISE} W4 channel image ($\lambda_{\rm iso} \sim 22~\micron$),
we use the 24 \micron\ band image of
the Multiband Imaging Photometer for {\it Spitzer} \citep[MIPS:][]{rie04}
obtained in 2007 January (AORKEY: 17657088,
17657344)\footnote{\url{http://sha.ipac.caltech.edu/applications/Spitzer/SHA/}},
which provides better spatial resolution ($6\arcsec$) than {\it WISE} W4.
The data used here are the Post-BCD image (``Level 2''),
which is automatically processed by the MIPS software pipeline
(version: S18.12.0).
As in the {\it WISE} data, zodiacal light is one of the significant sources of
background radiation in this waveband.
We therefore subtracted the background brightness from the mosaic image
using the model calculation \citep{kel98}.

Recent FIR images of the remnant were observed with PACS and SPIRE
on board {\it Herschel Space Observatory} \citep{pil10}
in 2009 September and December.
The two instruments have six photometric bands 
(70, 100, and 160 \micron\ for PACS, and
250, 350, and 500 \micron\ for SPIRE),
and they provide spatial resolutions of
$5\farcs2$, $7\farcs7$, $12\arcsec$, $18\arcsec$, $25\arcsec$, and $37\arcsec$,
respectively.
The fully calibrated data sets were downloaded from the {\it Herschel} Science
Archive\footnote{\url{http://www.cosmos.esa.int/web/herschel/science-archive/}},
and we use the data product levels of 2.5 and 2.0
for PACS and SPIRE, respectively.
All imaging data have well-established zero-level brightness,
whereas the PACS photometer images at level 2.5 do not
(PACS Observer's Manual - version 2.5.1).
To obtain the true zero point of the images,
we calibrated the zero level by comparison with
reference images whose zero point is well established.
For the reference photometric calibrator,
we used 60 and 100 \micron\ images
from Improved Reprocessing of the {\it IRAS} Survey \citep[IRIS:][]{miv05}
together with DIRBE 140 and 240 \micron\ and
{\it Planck} 350 and 550 \micron\ maps.
In Cas A, synchrotron radiation is a dominant source in the FIR waveband.
This nonthermal component is subtracted by
using the VLA radio image at 6 cm \citep{del04}.
The expected brightness at each isophotal wavelength 
is extrapolated from the power law 
$S_\nu = C~\nu^\alpha$ (erg cm$^{-2}$ s$^{-1}$ Hz$^{-1}$).
In this paper, we adopt $\alpha=-0.682$ and $C=2.50\times10^9$
erg cm$^{-2}$ s$^{-1}$ Hz$^{-1}$,
which are derived from the total brightness of the remnant
at 30, 44, 70, and 100 GHz using {\it Planck} \citep{pla11} archival
data\footnote{\url{http://irsa.ipac.caltech.edu/applications/planck/}}.

%%%%%%%%%%%%%%%%%%%%%%%%%%%%%%%%%%%%%%%%%%%%%%%%%%%%%%%%%%%%%%%%%%%%%%%%%%%80==>
\section{NIR Extinction by SN Dust} \label{sec-ext}
%%%%%%%%%%%%%%%%%%%%%%%%%%%%%%%%%%%%%%%%%%%%%%%%%%%%%%%%%%%%%%%%%%%%%%%%%%%80==>
\subsection{NIR Extinction Measurement} \label{sec-ext-nir}

We derive the selective extinction or ``color excess'' between 
1.26 and 1.64 \micron, $E(J-H)$, toward the infrared knots from
\begin{equation}
E(J-H)\equiv A_{1.26}-A_{1.64} 
= 1.086 ~ \ln \dfrac {[F_{1.26}/F_{1.64}]_{\rm int}}
{[F_{1.26}/F_{1.64}]_{\rm obs}}, 
\label{eq-01}
\end{equation}
where $A_{1.26}$ and $A_{1.64}$ are the 
total extinctions at 1.26 and 1.64 \micron, 
and $[F_{1.26}/F_{1.64}]_{\rm obs}$ and
$[F_{1.26}/F_{1.64}]_{\rm int}$ are the observed and intrinsic flux ratios
of the \feii\ 1.26 and 1.64 \micron\ lines, respectively.
These two strong \feii\ lines share the same upper state ($a^{4}D_{7/2}$),
so their intrinsic ratio is fixed by
their Einstein $A$ coefficients and wavelengths,
i.e., $[F_{1.26}/F_{1.64}]_{\rm int} =(A_{ki,1.26}/1.26)/(A_{ki,1.64}/1.64)$. 
The ratio has been derived from both theoretical calculations and observations,
and it ranges from 0.94 to 1.49 in previous studies
\citep[e.g.,][and references therein]{gia15,koo15}.
This large uncertainty of the intrinsic flux ratio hampers
accurate measurement of the absolute extinction toward the remnant
(this issue will be addressed in Section~\ref{sec-ext-sel}).
It is, however, worth noting that
the uncertainty does not affect the difference in $E(J-H)$ (Equation~\ref{eq-01})
among the knots, which gives the relative extinction to them.
In this paper, we adopt the intrinsic line flux ratio of 1.36, 
which is the theoretical value proposed by \citet{nus88} and \citet{deb10}.

%%%%%%%%%%%%%%%%%%%%%%%%%%%%%%%%%%%%%%%%%%%%%%%%%%%%%%%%%%%%%%%%%%%%%%%%%%%80==>
\subsection{Total Extinction to Cas A} \label{sec-ext-ext}

\begin{figure}
\center{
\includegraphics[scale=0.9]{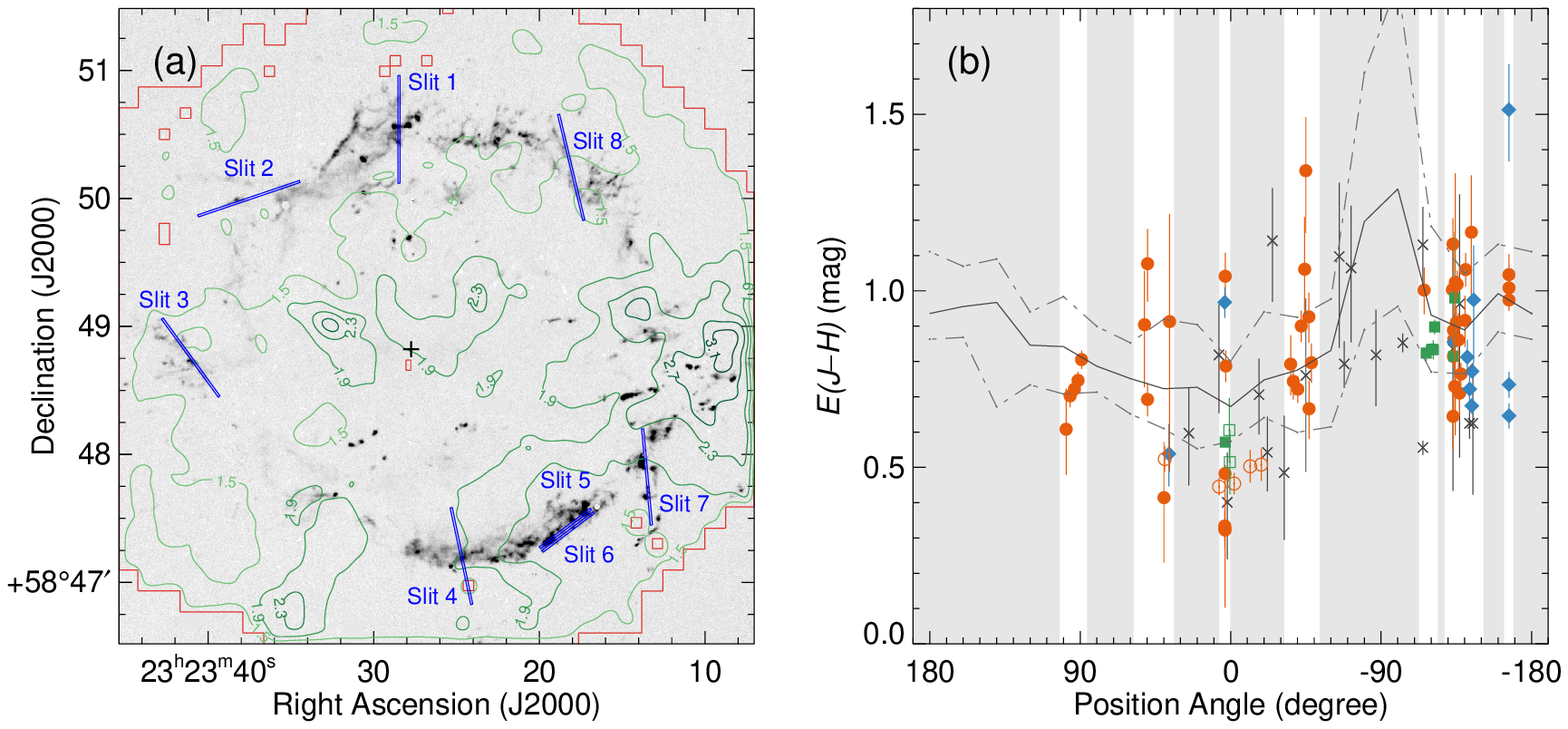}
}
\caption{
   (a) Slit positions (blue bars) marked on the continuum-subtracted
   \feii\ 1.64 \micron\ narrow-band image obtained in 2008 August.
   The black cross mark corresponds to the center of expansion \citep{tho01}.
   The green contours represent the hydrogen column density ($N_{\rm H}$)
   derived from a single-component fit of {\it Chandra} X-ray data
   \citep{hwa12}, and the levels are 1.5, 1.9, 2.3, 2.7, and
   3.1 $\times ~10^{22}~{\rm cm}^{-2}$.
   The red solid line is the boundary of the $N_{\rm H}$ map.
   (b) Variation of $E(J-H)$ with position angle.
   The position angle is the angle along the perimeter of the remnant
   measured from north to east with $PA=0$ corresponding to
   a line drawn due north from the expansion center (black cross mark)
   of Cas A.
   The filled symbols represent the 63 infrared knots detected \citep{koo13},
   with green squares indicating He-rich knots,
   red circles indicating S-rich knots, and
   blue diamonds indicating Fe-rich knots.
   The seven optical knots of \citet{hur96} are marked by open symbols: 
   five FMKs and two QSFs appear as red circles and green squares, respectively.
   The 19 infrared knots observed in \citet{eri09} are
   also marked by black X symbols.
   The gray-shaded regions are the areas uncovered in our observation.
   The black solid line shows the variation of the average $E(J-H)$
   along the iron-bright rim obtained from the $N_{\rm H}$ map of
   \citet{hwa12},
   and the black dashed-dotted lines represent maximum and minimum $E(J-H)$
   at each position angle (see the text for details).
} \label{fig-slit}
\end{figure}

In Figure~\ref{fig-slit}(b), 
we plot $E(J-H)$ for the 63 infrared knots as a function of position angle.
The colors and symbols represent the characteristic groups
classified by \citet{koo13}:
helium-rich knots in green squares, sulfur-rich knots in red circles, 
and iron-rich knots in blue diamonds.
In short, 
helium-rich knots are slowly moving ($\lesssim 100~\kms$) knots 
with strong \hei\ lines, the properties of which match those of
``quasi-stationary flocculi'' (QSFs), the circumstellar material
swept up by an SN blast wave \citep{van71,lee14}.
Sulfur-rich and iron-rich knots are fast-moving ($\gtrsim 100~\kms$) knots
with strong \sii\ and strong \feii\ lines, respectively.
They are SN ejecta material synthesized in different layers of the SN
and correspond to fast-moving knots (FMKs) in previous optical studies
\citep[e.g.,][]{van71,ham08}. 
For more detailed explanations on
the spectroscopic and kinematic properties of these knots, 
please refer to \citet{koo13}.
For comparison, we also plot $E(J-H)$ for the optical and infrared knots
measured in previous studies.
\citet{hur96} observed optical spectra of five FMKs and two QSFs
located in the northern bright rim and
measured $E(B-V)$ using ratios of the \sii\ and the Balmer lines, respectively.
We converted it to $E(J-H)(=0.30 \times E(B-V))$ 
assuming the general interstellar dust composition with
$R_{V}$ of 3.1 \citep{dra03},
and the open red circles and green squares in the figure represent
the FMKs and QSFs, respectively.
\citet{eri09} performed NIR spectroscopy for four slit positions
and derived the line ratio of \feii\ 1.26--1.64 \micron\ for 19 infrared knots
without any classification of their origin.
We obtained $E(J-H)$ using Equation~\ref{eq-01} and
marked the results by black X marks.

The derived $E(B-V)$ varies considerably over the remnant.
It appears that the extinction toward the west is 
systematically larger than that toward the east,
though there is a large scatter.
This systematic variation
is well known from previous radio and X-ray observations
\citep[e.g.,][]{keo96,rey02,hwa12}.
For example, the green contours in 
Figure~\ref{fig-slit}(a) show the column density ($N_{\rm H}$) map 
of \citet{hwa12} obtained from an analysis of the {\it Chandra} X-ray data. 
The column density is large ((2--3) $\times 10^{22}~{\rm cm}^{-2}$) toward 
the west and southwest directions, and 
molecular line studies showed that it is due to 
molecular clouds located in the Perseus Spiral Arm. 

Figure~\ref{fig-slit}(b) shows that the extinction 
varies considerably {\em within a slit}. For example, along Slit 4, 
$E(J-H)$ varies from 0.6 to 1.5 mag even though  
the emission is from a thin filament. This variation is significantly 
larger than what we would expect from the variation in the 
foreground extinction. 
For example,
the solid line in the figure represents an expected $E(J-H)$ variation
along the ejecta shell derived from the $N_{\rm H}$ map
in Figure~\ref{fig-slit}(a).
The ejecta shell is determined in the continuum-subtracted \feii\ image 
where the brightness is greater than 3$\sigma$ of the background rms noise, 
and the mean $N_{\rm H}$ is obtained at every $10\degr$ in position angle.
The column density is converted to $E(J-H)$ 
using 
$N_{\rm H}/E(B-V) = 5.8 \times 10^{21}~{\rm atoms~cm^{-2}~mag^{-1}}$
\citep{boh78}
and 
$E(J-H) = 0.30 \times E(B-V) = N_{\rm H}/1.9 \times 10^{22}~{\rm cm}^{-2}$.
The dashed-dotted lines represent the maximum and minimum $E(J-H)$
at each position angle.
It is obvious that, in several slits, 
the variation of $E(J-H)$ within a slit is much larger than the range
allowed from the $N_{\rm H}$ map.
As we show in the next section, this large variation 
is due to extinction {\em within the SN ejecta}.

%%%%%%%%%%%%%%%%%%%%%%%%%%%%%%%%%%%%%%%%%%%%%%%%%%%%%%%%%%%%%%%%%%%%%%%%%%%80==>
\subsection{Self-extinction within Cas A} \label{sec-ext-sel}

An interesting correlation is found when 
we plot $E(J-H)$ as a function of radial velocity, which is 
shown in Figure~\ref{fig-radvel}(a).
It is clear that 
the redshifted knots are generally more obscured than the blueshifted knots.
Because we are observing an expanding shell,
this suggests that the knots on the far side experience more extinction
than those on the front side, or that 
there could be extinction originating {\em within} the remnant.
To confirm this ``self-extinction,''
it is necessary to remove the spatially varying foreground interstellar extinction.
We subtract the extinction
derived from the X-ray observation (Figure~\ref{fig-slit}),
and the result is shown in Figure~\ref{fig-radvel}(b).
The correlation becomes slightly weaker,
but it is still obvious that
the redshifted knots are more heavily obscured than the blueshifted knots.
Therefore, our result implies that there is extinction,
which might be due to newly formed SN dust within the remnant.
The large scatter in extinction for the same velocity knots
might reflect differences among different sight lines,
i.e., non-uniform distribution of SN dust.

Note in Figure~\ref{fig-radvel}(b) that
the minimum $E(J-H) \sim -0.3$, 
which means that the $E(J-H)$ derived from the $N_{\rm H}$ map is 
higher than the $E(J-H)$ derived from \feii\ line ratios.
We can think of two possibilities for the negative $E(J-H)$.
First, it could be due to the uncertainty of the theoretical \feii\ line ratio.
To compensate for the $E(J-H)$ of $-0.3$ mag, however,
the theoretical flux ratio of \feii\ 1.26--1.64 \micron\
should be more than 1.8, which is considerably higher than the numerical value 
obtained in previous studies ($0.94 \le [F_{1.26}/F_{1.64}]_{int} \le 1.49$).
Second, the $N_{\rm H}$ derived from X-ray data could be overestimated.
Indeed, the mean column density from the X-ray studies
is $\sim 1.5 \times 10^{22}~{\rm cm}^{-2}$ 
\citep{wil02, hwa12}, which is 
higher than those of radio observations, e.g., 
$\sim 1.1 \times 10^{22}~{\rm cm}^{-2}$ \citep{tro85,keo96}.
We also note that the column densities of \cite{hwa12} 
are higher by (0.4--0.7) $\times 10^{22}$~cm$^{-2}$
than those of \cite{lee14}, who carefully analyzed
the X-ray spectra around the outer SNR shock.
The difference of 0.3 mag in $E(J-H)$, which corresponds 
to $N_{\rm H} \approx  6 \times 10^{21}~{\rm cm}^{-2}$,
may therefore have resulted from the  
difficulty in background removal in the analysis of \cite{hwa12}.
Another complication is that 
the X-ray-absorbing column also includes self-extinction, so the 
X-ray-based extinction overestimates the foreground extinction.
Thus it appears difficult to obtain an accurate map of the  
foreground extinction from the X-ray data alone, but this uncertainty 
is not likely to erase  
the systematic correlation in Figure~\ref{fig-radvel}(b).  

The non-uniform spatial distribution of SN dust and the 
uncertainty in the foreground extinction 
hamper the analysis of the correlation in Figure~\ref{fig-radvel}(b).
However, in one slit (Slit 4), we detected convincing evidence 
for self-extinction without those sight-line-dependent complications.
Figures~\ref{fig-pv5}(a) and (b) show that
the velocity structure of the \feii-line-emitting gas in Slit 4 is ``arc''-like,
suggesting that the slit crosses a portion of an expanding shell.
This is consistent with the result of previous studies 
that the main ejecta shell is a thin shell
expanding at 4000--5000 $\kms$ \citep[e.g.,][]{del10,ise12}.
In Slit 4, therefore, 
we have both blue and redshifted Fe knots along a given sight line,
and the uncertainty due to different sight lines disappears.
Now Figure~\ref{fig-pv5}(c) shows the $E(J-H)$ map
obtained for pixels with an \feii\ 1.64 \micron\ brightness greater than 
3$\sigma$ above the background noise.
A systematic increase in the extinction 
toward the redshifted knots is apparent. 
We computed an intensity-weighted $E(J-H)$ as a function of radial velocity,
which is shown by the black solid line in Figure~\ref{fig-pv5}(d). 
We also divided the iron filament into three subregions
along the slit (A--C in Figure~\ref{fig-pv5}(c)),
and their mean $E(J-H)$ values are over-plotted on the figure
with red, green, and blue solid lines.
The total $E(J-H)$ within the ejecta shell from Figure~\ref{fig-pv5}(d) 
is $0.23\pm0.05$ mag with a systematic gradient of 
$0.13\pm0.02$ mag per 1000 km s$^{-1}$.
This should be mostly, if not entirely, due to SN dust
because the blue and redshifted ejecta materials in the slit
are essentially along the same sight line.
There are small-scale variations in $E(J-H)$ in Figure~\ref{fig-pv5}(d). 
They could be due to either incomplete subtraction of 
OH airglow emissions (e.g., gray-hatched areas in the figure) or
very small dust clumps within the slit, or both.

The obtained $E(J-H)$ can be converted to
the dust column density ($\Sigma_{\rm d}$) with an appropriate 
$\Delta \kappa_{JH}\equiv \kappa_{1.26}-\kappa_{1.64}$ (cm$^{2}$ g$^{-1}$),
where $\kappa_{1.26}$ and $\kappa_{1.64}$ are 
the mass absorption coefficients at 
1.26 and 1.64 \micron, respectively:
\begin{equation}
E(J-H) = 1.086 \int \Delta \kappa_{JH}~\rho_{\rm d}~dl 
       \approx 0.11 
\left( \dfrac{\Delta \kappa_{JH}}{10^3~{\rm cm^2~g^{-1}}} \right)
\left( \dfrac{\Sigma_{\rm d}}{10^{-4}~{\rm g~cm^{-2}}} \right)
\label{eq-02}
\end{equation}
This equation yields the column density of a given dust species
required to explain the observed NIR self-extinction.
For example, if it is Fe dust, which has
$\Delta \kappa_{JH} \sim 10^{3}$ cm$^{2}$ g$^{-1}$
\citep[][see Figure~\ref{fig-size}]{sem03},
the required dust column is $\sim 10^{-4}$ g cm$^{-2}$.

%%%%%%%%%%%%%%%%%%%%%%%%%%%%%%%%%%%%%%%%%%%%%%%%%%%%%%%%%%%%%%%%%%%%%%%%%%%80==>
\section{FIR Emission and SN Dust Composition} \label{sec-sed}
%%%%%%%%%%%%%%%%%%%%%%%%%%%%%%%%%%%%%%%%%%%%%%%%%%%%%%%%%%%%%%%%%%%%%%%%%%%80==>
\subsection{FIR Emission from SN Dust} \label{sec-sed-fir}

What dust species would
produce the $E(J-H)$ of 0.2 mag detected in Slit 4?
In this section, we constrain the dust species by investigating the 
FIR emission properties of SN dust.

\begin{figure}
\center{
\includegraphics[scale=0.9]{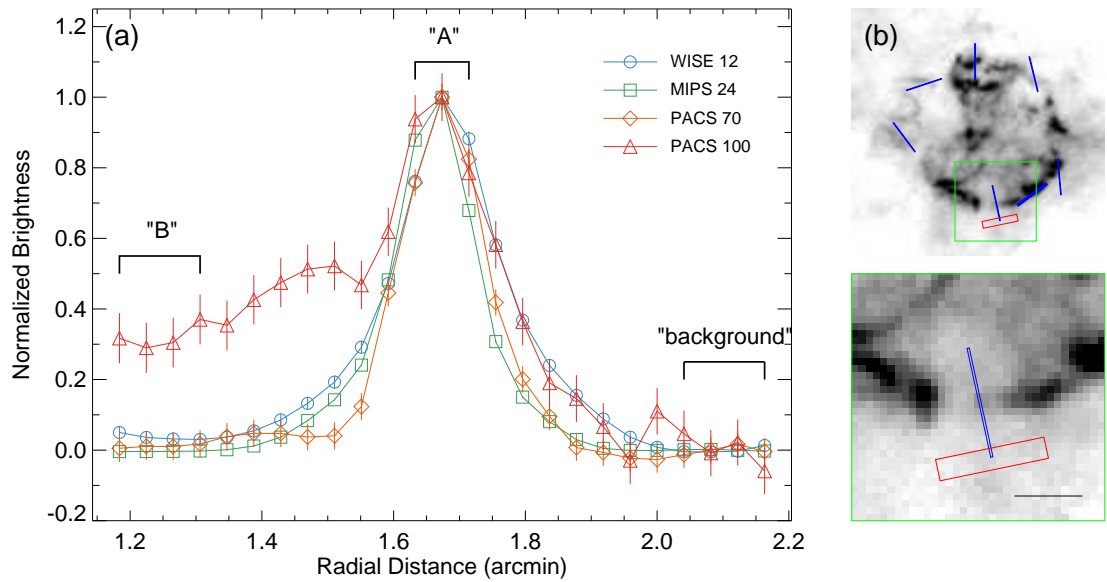}
}
\caption{
   (a) One-dimensional profiles of 12, 24, 70, and 100 \micron\
   brightnesses along the length of Slit 4.
   Each profile is background-subtracted
   and normalized by the peak brightness at $1\farcm67$.
   (b) Location of the background region (red box; $10\arcsec \times 50\arcsec$
   between $2\farcm0$ and $2\farcm2$ in radial distance)
   on the synchrotron-subtracted PACS 100 \micron\ image.
   Slit positions are also marked with blue boxes.
   The bottom image is a zoomed-in image of the area
   in the green box in the top frame,
   and the black scale bar in the lower right represents
   an angular scale of $30\arcsec$.
} \label{fig-profile}
\end{figure}

Figure~\ref{fig-profile}(a) shows the one-dimensional (1D) brightness profiles
at 12, 24, 70, and 100 \micron\ along the slit length in Slit 4. 
Note that these data 
have similar angular resolutions of $5\farcs2$--$7\farcs7$
(Section~\ref{sec-obs-mir}).
The profiles are background-subtracted using the mean brightnesses of 
an area just outside the ejecta shell
(red square in Figure~\ref{fig-profile}(b))
and are normalized by their peak brightnesses.
All four profiles have maxima at $1\farcm67$ 
where the iron filament is located (Figure~\ref{fig-pv5}).
All emission drops to zero inside of the ejecta shell except for
the 100 \micron\ emission, which remains constant at 
$\sim30$\% of its peak brightness.
This excess emission at 100 \micron\ must be mostly from the 
unshocked cool SN dust detected previously in the FIR 
\citep{bar10,are14}.

We examine the SEDs at two positions (see Figure~\ref{fig-profile}(a)):
(1) the peak brightness position at $1\farcm67$
with $\sim 6\arcsec$ aperture (hereafter position A) and
(2) the inner 100 \micron\ excess region between $1\farcm2$ and $1\farcm4$
(hereafter position B).
For comparison, we also derive an interstellar dust SED 
just outside of the shell (red square in Figure~\ref{fig-profile}(b))
where the radiation from the remnant is almost negligible.
To derive accurate parameters for the SN dust,
we first have to subtract the contributions of line emissions 
from the measured brightnesses.
For this purpose,
we retrieved the archival MIR and FIR spectra from {\it Spitzer}
IRS\footnote{http://sha.ipac.caltech.edu/applications/Spitzer/SHA/}
and {\it ISO} LWS\footnote{http://iso.esac.esa.int/ida/}
observations, respectively.
We produced an IRS spectral data cube using CUBISM \citep{smi07} and 
extracted 1D spectra within a $6\arcsec$ aperture at positions A and B as 
well as the background position.
The strong emission lines
falling in the {\it WISE} W3 and MIPS 24 \micron\ bands are
the \arii\ 8.99 \micron, \siv\ 10.5 \micron, \neii\ 12.8 \micron,
\neiii\ 15.6 \micron, and \oiv+\feii\ 25.9 \micron\ lines,
and their contributions to the fluxes in the two bands have been estimated as
8.9\% and 4.8\% for position A, 4.6\% and 16\% for position B, 
and 2.0\% and 1.8\% for the background position, respectively. 
There are no corresponding FIR spectroscopic data,
but an {\it ISO} LWS 1D spectrum (43--190 \micron) 
from an area close ($\sim50\arcsec$) to Slit 4 is available
(\#3 spectrum of \citet{doc10}). 
According to this {\it ISO} spectrum,
the strong emission lines falling in the PACS 70 and 100 \micron\ bands are  
the \oi\ 63.2 and \oiii\ 88.4 \micron\ lines, respectively. 
Estimation of the contribution of these lines to the observed 
surface brightnesses at our positions is tricky because the {\it ISO} LWS 
beam is large, i.e., $\sim80\arcsec$ \citep{llo03},
and covers both the Cas A ejecta shell and the background area.
We estimated the line contribution by assuming that
both emission lines come from the ejecta shell,
not from the interstellar material.
This assumption is plausible because
the two lines are detected at several positions along the ejecta shell
to show similar broad profiles,
whereas there is little emission outside of the remnant \citep{doc10}.
Under this assumption, the two lines contribute only to the brightness of A,
which is estimated as 3.8\% and 4.5\% in the PACS 70 and 100 \micron\ bands,
respectively.
In the PACS 160 \micron\ band,
there is the \oi\ 146 \micron\ line, but its contribution is 
almost negligible. The observed and line-subtracted dust continuum
brightnesses at the three positions are listed in Table~\ref{tab-sed}. 

Figure~\ref{fig-sed} shows the line-emission-subtracted SEDs at 
positions A and B in red and blue solid lines, respectively. 
Both SEDs have broad thermal emission bumps peaking at 160 \micron,
which arise from the line-of-sight interstellar dust component at 
a temperature of $\sim 20$ K. This FIR SED matches well 
the SED of the background area outside the shell (black solid line), 
which represents the interstellar dust SED.
The background-subtracted SEDs,
which presumably represent ``pure'' SN dust SEDs,  
are shown in red and blue dotted lines in Figure~\ref{fig-sed}.
The SED at A now shows maximum brightness at 24 \micron,
and the brightness decreases toward longer wavelengths.
On the other hand, that at B peaks at 100 \micron, 
which implies that the dust temperature is much lower than that at A.
These two SEDs should represent the SEDs of shocked (+unshocked) 
and unshocked dust species, respectively. 
We will analyze the SED at position A, where we obtained 
self-extinction of $E(J-H)=0.23\pm0.05$. 

Before doing a detailed analysis in the next section, we 
first show that the shocked warm dust component emitting the MIR emission  
cannot explain the NIR color excess that we detected.
Previous observations showed that the warm dust is silicate grains 
\citep[e.g., ][]{are14}, and we assume that it is
MgSiO$_{3}$ grains with a size of 0.001--0.1 \micron, which has
$10^{-1}\lesssim \Delta \kappa_{JH} \lesssim 10^{2}$ cm$^2$ g$^{-1}$
\citep{dor95}. 
(As we will show in the next section,
assuming other silicate dust species will yield the same conclusion.) 
The SED at A has excess emission in the FIR,
and we perform a two-temperature modified blackbody fit
assuming that the cool dust has the optical properties of 
general interstellar dust \citep{dra03}, as in \citet{sib10}.
The surface brightness of each component at temperature $T$
is calculated from  
\begin{equation}
{S_{\nu}}=\kappa_{{\rm abs},\nu}~B_{\nu}(T)~\Sigma_{\rm d}
\label{eq-03}
\end{equation}
where $\kappa_{{\rm abs},\nu}$ is the mass absorption coefficient, and 
$B_{\nu}(T)$ is the Planck function.
We used MPFIT \citep{mar09},
which is a nonlinear least-squares fitting routine within IDL,
to derive the values of $T$ and $\Sigma_{\rm d}$ at which $\chi^2$ is minimum
as well as their formal 1$\sigma$ uncertainties.
The SED is well fitted with warm dust at $T = 115 \pm 2$ K plus
cool dust at $T = 40 \pm 6$ K (Figure~\ref{fig-sed}).
From this two-component fitting,
we found that almost all of the brightnesses at 12 and 24 \micron\
arise from the warm dust component,
whereas more than 80\% of the brightness in the FIR
longer than 100 \micron\ is from the cool dust component.
The warm and cool dust components yield
$\Sigma_{\rm d}=3.2 (\pm 0.3) \times 10^{-8}$ g cm$^{-2}$
and $1.1 (\pm 0.8) \times 10^{-6}$ g cm$^{-2}$, respectively.
From Equation~\ref{eq-02}, therefore,
the corresponding $E(J-H)$ of the warm dust with a size of 0.001--0.1 \micron\
is $10^{-8}-10^{-5}$ mag,
and that of the cool dust is $\sim0.003$ mag.
The column density of warm dust
required to explain the MIR emission is very small,
and the resulting NIR extinction is several orders of magnitude smaller than
the value we derived ($0.23 \pm 0.05$ mag). 
Therefore, it must be the cool dust component that is responsible 
for the NIR extinction. 
(Note that the grain species of the warm dust component and therefore
its contribution to the NIR extinction are rather well constrained.)  
However, the above result suggests that the cool dust component 
cannot have the optical properties of the general interstellar dust
because then the NIR extinction is again two orders of magnitude less than
the value we derived.
The grain species of the cool dust should have
large $\Delta \kappa_{JH}$ and/or
small FIR opacity ($\kappa_{{\rm abs},\nu}$) compared to
that of general interstellar dust.

%%%%%%%%%%%%%%%%%%%%%%%%%%%%%%%%%%%%%%%%%%%%%%%%%%%%%%%%%%%%%%%%%%%%%%%%%%%80==>
\subsection{SN Dust Composition} \label{sec-sed-com}
%%%%%%%%%%%%%%%%%%%%%%%%%%%%%%%%%%%%%%%%%%%%%%%%%%%%%%%%%%%%%%%%%%%%%%%%%%%80==>
\subsubsection{Dust Species and Their Optical Properties} \label{sec-sed-com-opt}

For the composition of SN dust,
we consider the grain species predicted by previous theoretical models
for CCSNe \citep{tod01,noz03,noz08,noz10}.
The grain species considered in this paper are summarized in Table 2 with
references for their optical constants ($n$ and $k$).
Note that carbon grains are formed in the He layer,
silicates (MgSiO$_3$, Mg$_2$SiO$_4$, and SiO$_2$) and
oxides (Al$_2$O$_3$ and MgO) in the O-rich layer,
and other heavy-element (Si, Fe, FeS) grains in the Si--S--Fe layer.
In addition, various types of oxide grains
(SiO$_2$, MgSiO$_3$, Mg$_2$SiO$_4$, Al$_2$O$_3$, and Fe$_3$O$_4$)
are produced in the mixed SN models.
According to \citet{rho08}, the MIR spectra of Cas A are well
fitted by the combination of these grain opacity curves,
which implies that
these are the major dust species produced by the SN explosion.
We calculated the absorption and extinction coefficients
($\kappa_{{\rm abs},\nu}$ and $\kappa_{{\rm ext},\nu}$)
using the Mie theory \citep{boh83}
assuming a spherical grain with a radius of 0.001, 0.01, or 0.1 \micron.

Figure~\ref{fig-grain} shows
$\Delta \kappa_{JH}$ and $\kappa_{{\rm abs},\nu}$
at 24 and 70 \micron\ for the grain species with different sizes.
As expected,
the $\kappa_{{\rm abs},\nu}$ values of dust species at 24 and 70 \micron\
do not depend on the grain size except for Fe dust.
For Fe dust, which is a metallic dust,
not only the electric dipole but also the magnetic dipole is important
for grains larger than 0.006 \micron,
the efficiency ($Q_{\rm abs}$) of which is proportional to the volume
and not the size \citep{tan84}.
Therefore, the FIR opacity 
($\kappa_{{\rm abs},\nu} = (\pi a^2 Q_{{\rm abs},\nu}) / (4/3 \pi a^3 \rho) $,
where $Q_{{\rm abs},\nu}$ is the absorption efficiency)
of Fe dust is proportional to the square of the size for large grains. 
Note that small Fe dust ($\le 0.01~\micron$) has an FIR dust opacity
much smaller than those of the other dust grains, e.g., 
$\le 0.2$ cm$^2$ g$^{-1}$ versus 
2--100 cm$^2$ g$^{-1}$ at 70 \micron.
$\Delta\kappa_{JH}$ in general does not depend on the grain size for 
$a \le 0.01~\micron$ and increases with $a$ for larger grains
up to $a\gtrsim 0.1~\micron$ (Figure~\ref{fig-size}).
This is the result of standard grain extinction properties:
for small grains, i.e., $x \equiv 2 \pi a/ \lambda \ll 1$,
$Q_{\rm ext}$($=Q_{\rm abs}+Q_{\rm sca}$) is dominated by 
$Q_{\rm abs}$, which is proportional to $x$,
so $\Delta \kappa_{JH}$($\varpropto (Q_{{\rm ext},J}-Q_{{\rm ext},H})/a$)
is independent of the grain radius
(except for Fe dust, which has $Q_{\rm abs} \varpropto a^3$).
For larger grains up to $x \lesssim 1$,
the scattering with $Q_{\rm sca} \varpropto x^4$ is more important,
so $\Delta \kappa_{JH}$ increases.
Because, in general, $Q_{\rm ext}$ increases to
a value of $Q_{\rm ext} \approx 3-5$ near $|m-1| x \approx 2$,
where $m (\equiv n+ik)$ is the complex refractive index,
and then converges to 2 \citep{dra11},
$\Delta \kappa_{JH}$ reaches a maximum near $|m-1| x \approx 2$
and then rapidly drops to zero.
For grains of $a=0.1~\micron$,
$\Delta\kappa_{JH}=10^2$--$10^{4}$ cm$^2$ g$^{-1}$.
Fe grains have $\Delta\kappa_{JH}\sim 10^3$ cm$^2$ g$^{-1}$
almost independent of their sizes.
As discussed in the previous section,
the dust grains responsible for the NIR extinction should have
large $\Delta \kappa_{JH}$ and/or small FIR $\kappa_{{\rm abs},\nu}$
compared to the general interstellar dust composition.
In this sense, small Fe (or other metallic) dust grains
seem to be a candidate for the cool dust.

%%%%%%%%%%%%%%%%%%%%%%%%%%%%%%%%%%%%%%%%%%%%%%%%%%%%%%%%%%%%%%%%%%%%%%%%%%%80==>
\subsubsection{Spectral Modeling} \label{sec-sed-com-mod}

We now perform spectral fitting to the SN dust SED at position A.
We assume that the SN dust is composed of two temperature components,
i.e., warm and cool dust, as noted in Section~\ref{sec-sed-fir}.
A single component cannot explain both the MIR peak and the FIR excess.
We found that essentially all of the MIR emission is from 
the warm dust. Therefore, we simply assume that
the 12 and 24 \micron\ brightnesses are from the 
warm dust, which then yields the 
temperature and surface density of warm dust without fitting the entire SED.
Color correction was made using the filter response curve
at each band and the spectral shape of each grain model.

\begin{figure}
\center{
\includegraphics[scale=0.9]{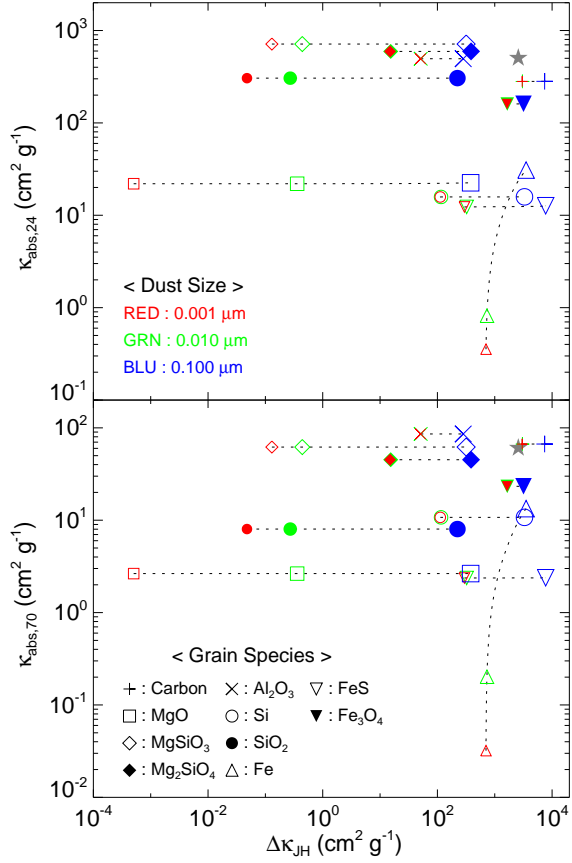}
}
\caption{
   $\kappa_{{\rm abs},\nu}$ at
   24 \micron\ ($\kappa_{{\rm abs},24}$; upper panel) and
   70 \micron\ ($\kappa_{{\rm abs},70}$; lower panel)
   versus $\Delta \kappa_{JH}$
   for 10 grain species in Table~\ref{tab-grain}. 
   Symbols and colors denote the grain species and sizes, respectively.
   We connect the same symbols with black dotted lines.
   For comparison,  the general interstellar dust \citep{dra03} is marked with a star symbol.
} \label{fig-grain}
\end{figure}

\begin{figure}
\center{
\includegraphics[scale=0.7]{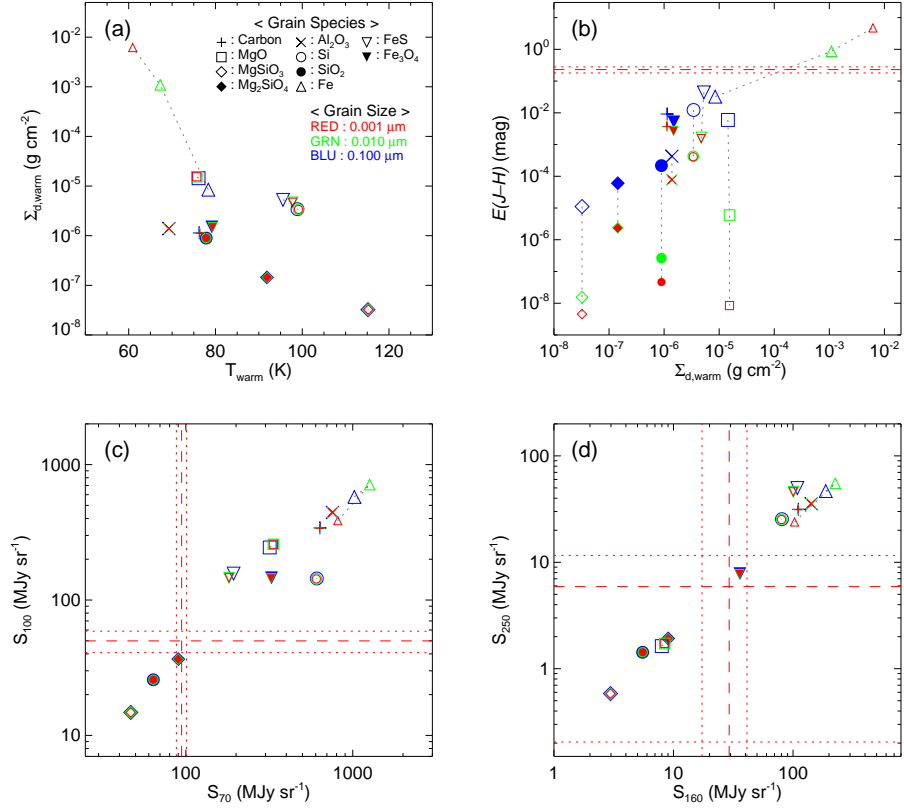}
}
\caption{
   (a) Temperature and column density of warm dust grains obtained
       from the fit of MIR emission.
       Symbols and colors are the same as in Figure~\ref{fig-grain}.
   (b) Predicted $E(J-H)$ of warm dust grains from their column densities in (a). 
       The red dashed and dotted lines are the measured color excess and 
       its 1$\sigma$ uncertainty, respectively.
   (c) Expected brightnesses of warm dust grains at 70 and 100 \micron.
       The red dashed and dotted lines represent 
       the observed brightnesses and their 1$\sigma$ uncertainty, respectively.
   (d) Same as (c) but for 160 and 250 \micron\ brightnesses.
} \label{fig-warm}
\end{figure}

Figure~\ref{fig-warm}(a) shows the derived temperature and 
surface density of warm dust for different species.
Because the FIR $\kappa_{{\rm abs},\nu}$ values of dust grains,
except for Fe dust, do not depend on their sizes (Figure~\ref{fig-grain}),
the derived temperature and column density of dust species 
are independent of the grain size. 
Not all of these dust species can match the FIR part of the SED.
This is shown in Figures~\ref{fig-warm}(c) and (d), where we 
compare the expected FIR brightnesses to 
the observed brightnesses (the red dashed and dotted lines).
Note that it is acceptable if the expected brightness is  
less than the observed brightness because the difference can be contributed 
by cool dust, but not vice versa. 
Therefore, only three dust species can match the FIR part of the SED:
MgSiO$_{3}$, Mg$_{2}$SiO$_{4}$, and SiO$_{2}$.
The temperature and column density of the three grain models are
80 K $<T_{\rm warm}<$ 120 K and
$10^{-8}$ g cm$^{-2} < \Sigma_{\rm d,warm} < 10^{-6}$ g cm$^{-2}$,
respectively.
The compositions and corresponding temperatures of this warm dust component
are well consistent with previous observational and theoretical results
\citep{rho08,noz10}.
Figure~\ref{fig-warm}(b) shows  
the expected $E(J-H)$ of warm dust estimated from
their surface density using Equation~\ref{eq-02}.
Note that the $E(J-H)$ values of the three dust species  
are less than $\sim 10^{-4}$ mag,
which is a few orders of magnitude smaller than the observed $E(J-H)$,
as we already pointed out in Section~\ref{sec-sed-fir}.

\begin{figure}
\center{
\includegraphics[scale=0.7]{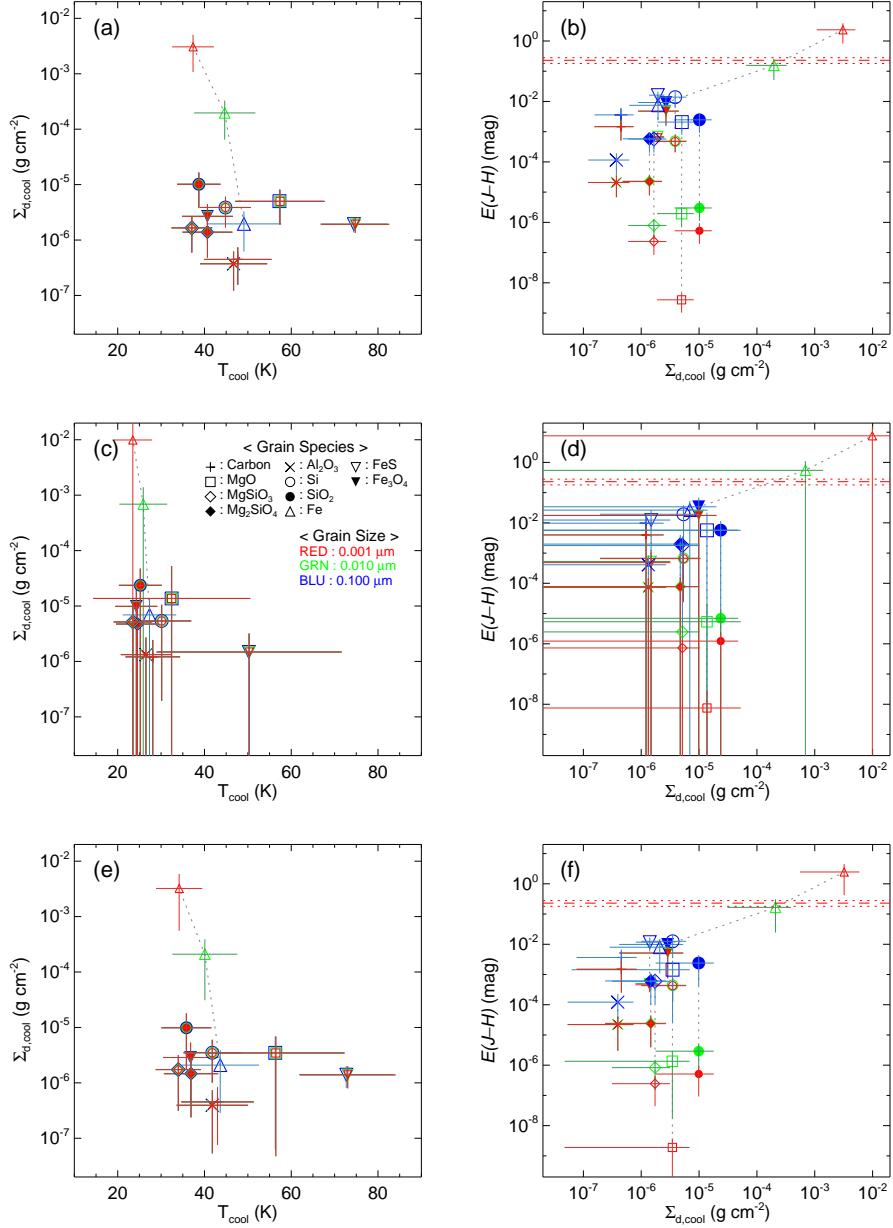}
}
\caption{
   (a) Temperature and column density of various cool dust species
       obtained from the fit of FIR emission
       assuming that the warm dust component is MgSiO$_3$.
   (b) Predicted $E(J-H)$ of cool dust from the column density in (a).
       The red dashed and dotted lines are the measured color excess and 
       its 1$\sigma$ uncertainty, respectively.
   (c)--(d) Same as (a)--(b)
       but assuming the warm dust component is Mg$_2$SiO$_4$.
   (e)--(f) Same as (a)--(b)
       but assuming the warm dust component is SiO$_2$.
} \label{fig-cool}
\end{figure}

With the warm dust component constrained,
we can now search for possible candidates for the cool dust component. 
We perform a two-temperature modified blackbody fit of the SED 
for each of the three possible warm dust species above,
the temperature and surface density of which are fixed.
The left column of Figure~\ref{fig-cool} shows 
the derived temperature and surface density of the cool dust 
for each warm dust species. 
All combinations can fit the observed SED 
within the 1$\sigma$ uncertainty of the measured brightness
except when the cool dust is MgO grains. 
The opacity of MgO has a small bump peaking at 100 \micron,
so it cannot properly fit the observed brightness
at either 100 or 160 \micron.
The temperature of the cool dust ranges from 
30 to 70 K depending on its composition.
When the warm dust is Mg$_2$SiO$_4$, the cool dust 
temperature is systematically lower.
The right column of Figure~\ref{fig-cool} shows the 
expected color excess of each cool dust species.
We see that small ($\leq 0.01~\micron$) Fe grains 
can match the observed $E(J-H) \approx 0.2$ mag for any warm dust species
with $\Sigma_{\rm d}=2-5\times 10^{-4}$~g cm$^{-2}$. 
None the other grain species can produce sufficient extinction. 

\begin{figure}
\center{
\includegraphics[scale=0.9]{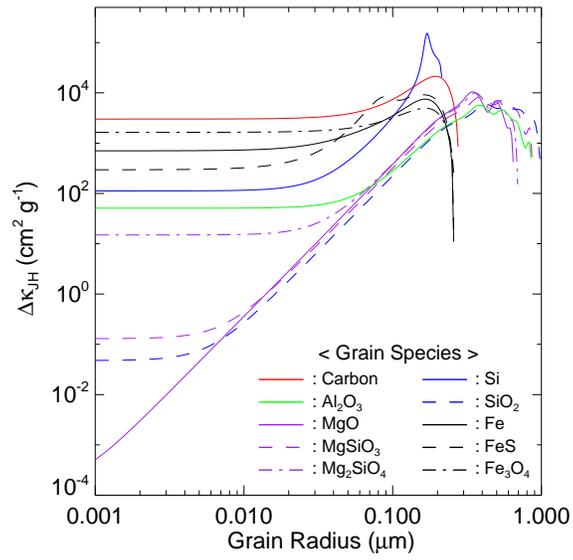}
}
\caption{
   $\Delta \kappa_{JH}$ versus grain radius for 10 grain species.
} \label{fig-size}
\end{figure}

Figure~\ref{fig-cool}, however, shows that as the grain size increases, 
$E(J-H)$ increases because larger grains have higher $\Delta \kappa_{JH}$
(Figure~\ref{fig-grain}),
so perhaps grains larger than 0.1 \micron\
might also produce $E(J-H) \approx 0.2$ mag.
Figure~\ref{fig-size} shows $\Delta \kappa_{JH}$ for each grain species
as a function of grain radius.
For oxide grains (i.e., MgO, MgSiO$_3$, Mg$_2$SiO$_4$, Al$_2$O$_3$, and SiO$_2$),
$\Delta \kappa_{JH}$ 
reaches a maximum at $\sim 0.4~\micron$ and
then drops abruptly at $\gtrsim 0.6~\micron$,
whereas for metallic grains (i.e., Si, Fe, FeS, and Fe$_3$O$_4$)
and amorphous carbon, the transition occurs at somewhat shorter wavelengths.
We find that pure Si grains with a radius of 0.16--0.21 \micron\
can also explain the color excess we observed
because of their large $\Delta \kappa_{JH}$ (Figure~\ref{fig-size}),
although the permitted size range appears to be too narrow.
Table~\ref{tab-param} summarizes the parameters of the warm and cold dust grains 
that can explain both the FIR SED and the NIR extinction.

%%%%%%%%%%%%%%%%%%%%%%%%%%%%%%%%%%%%%%%%%%%%%%%%%%%%%%%%%%%%%%%%%%%%%%%%%%%80==>
\section{Discussion} \label{sec-dis}
%%%%%%%%%%%%%%%%%%%%%%%%%%%%%%%%%%%%%%%%%%%%%%%%%%%%%%%%%%%%%%%%%%%%%%%%%%%80==>
\subsection{Cool SN Dust Responsible for the NIR Extinction} \label{sec-dis-coo}
 
According to our result, the cool dust that can explain
the observed NIR extinction in the southern part of Cas A (i.e., in Slit 4)
could be either 
small ($\lesssim 0.01~\micron$) Fe or large ($\gtrsim 0.1$ $\micron$) Si grains.
In this section, we discuss these two possibilities.

In the Slit 4 position, the reverse shock is now encountering pure Fe ejecta 
in the innermost region \citep{koo13}.
Considering this, 
cool SN dust mainly consisting of Fe grains seems to be reasonable.
The small size also appears to be consistent with the theoretical prediction that 
the average radius of newly formed SN dust grains in SN IIb
is smaller than 0.01 \micron\ because of low gas density in the 
expanding SN ejecta 
\citep{koz09,noz10}. 
On the other hand, theory also predicts that 
the gas density of the innermost ejecta 
is too low to produce pure Fe grains when the temperature of the region
drops to 800 K, at which Fe grains can start to coagulate,
so pure Fe dust is not expected to form in SNe IIb \citep{koz09,noz10}. 
However, the above conclusion assumes a spherically symmetric explosion with  
homogeneous and stratified ejecta.
If the gas density is inhomogeneous and/or the SN explosion is asymmetric,
we cannot rule out the possibility of pure Fe grain formation
in the innermost SN ejecta, where the gas density is relatively high
\citep{sar14}.
Indeed, the ejecta material emitting \feii\ lines is dense,
with electron densities of $\sim 10^4$~cm$^{-3}$ \citep{koo13}.

Another possible grain species responsible 
for the large $E(J-H) \approx 0.2$ mag is large Si grains.
This possibility may be supported by previous infrared studies;
{\it Spitzer} spectral mapping observations
found strong \silii\ emission together with \oiv\ and \siii\ lines 
in the interior of the remnant, which arises from the unshocked SN ejecta
photoionized by UV and X-ray emission from the Bright Rim
\citep[e.g.,][]{smi09,ise10,are14,mil15}.
The presence of ionized silicon in the interior
supports the suggestion that the cool dust could be Si and/or silicate grains.
Theoretically, pure Si grains can form in the Si-rich layer, 
but the grain size is very small, i.e., $\lesssim 0.01~\micron$ \citep{noz10}.
However, again if the SN ejecta is clumped,
grains larger than 0.1 \micron\ can form in dense clumps \citep{sar14},
so large Si grains are not implausible.

%%%%%%%%%%%%%%%%%%%%%%%%%%%%%%%%%%%%%%%%%%%%%%%%%%%%%%%%%%%%%%%%%%%%%%%%%%%80==>
\subsection{Composition of SN Dust in Cas A} \label{sec-dis-com}

From previous studies, it is reasonably well established that
there are two SN dust components in Cas A:
warm ($\sim 100$ K) dust swept up by the reverse shock and 
cool (30--40 K), unshocked dust residing in the interior of the remnant. 
The composition of the warm dust is relatively well specified from  
its MIR spectral features:
Mg-silicates, SiO$_2$, Al$_2$O$_3$, and/or C glass.
The estimated total mass of warm dust ranges from 0.008 to 0.054 $\msun$. 
The composition of the cool dust has been essentially unknown
because in the FIR, the SED is mostly smooth without prominent spectral features.
Assuming the FIR absorption coefficients of
the general interstellar dust \citep{sib10} or silicates \citep{bar10},
a total mass of $\sim0.07~\msun$ has been obtained.

Our analysis also showed that there are warm and cool dust components 
in the southern ejecta shell. The derived temperature (80--120 K) 
and composition (MgSiO$_3$, Mg$_2$SiO$_4$, or SiO$_2$)   
of the warm dust are consistent with previous results. 
For the cool dust, we were able to show for the first time that
its composition is either 
small Fe or large Si grains.
Their temperatures are 30--50 K.
These grains are not expected in uniformly expanding stratified SN ejecta,
but they can possibly form in dense clumps.
 
At first glance,
the different compositions of warm and cool dust appear to be incompatible
if we consider that the cool dust is swept up by the reverse shock
and subsequently turns into warm dust with a temperature of $\sim 100$ K.
However, this can be understood if we consider that the SN ejecta is clumped.
As we pointed out above,
the cool dust responsible for the NIR extinction resides in dense clumps. 
When the dense clumps are swept up by the reverse shock,
the shock speed is slow (e.g., $\lesssim 10~\kms$),
and the gas temperature is below $10^4$ K \citep{koo13},
so the dust grains in the clumps are not heated to high temperature.
On the other hand, when the diffuse ejecta is swept up by the reverse shock, 
the gas is heated to X-ray-emitting temperature
($\sim 10^7$~K), and the dust grains there are heated by collisions with 
electrons to high temperature. 
Therefore, it is the shock-heated warm grains in the diffuse ejecta
that emit most of the MIR and FIR ($\lesssim 70~\micron$) emission
(Figure~\ref{fig-sed}),
and their compositions could be different from those of grains in dense clumps.
In summary, our observation can be explained if 
the SN ejecta is clumped and
the grain species in diffuse material and dense clumps are different, 
i.e., silicate grains in the diffuse material and
small Fe or large Si grains in the dense clumps. 
The cool dust that we detected could reside
either in unshocked or shocked clumps,
and it is these cool grains that are responsible for the NIR extinction.

When the cool dust is pure Fe grains, 
we can make a crude estimation of  
the dust formation efficiency from our observation. 
The required Fe dust mass density of (2--5) $\times 10^{-4}$ g cm$^{-2}$
corresponds to an Fe nucleus column density of (2--5) $\times 10^{18}$ cm$^{-2}$.
For comparison, the characteristic size of \feii-line-emitting clumps 
is $5\arcsec$ or 0.08 pc at 3.4 kpc,
and their average electron density is $2\times 10^4$ cm$^{-3}$ \citep{koo13}.  
For pure Fe ejecta clumps, because Fe is mostly in Fe$^+$,
it means that the number density of Fe atoms is $2\times 10^4$ cm$^{-3}$,
and the characteristic Fe column density in the gas phase of Fe ejecta clumps  
is $N_{\rm Fe} \sim 5 \times 10^{21}$ cm$^{-2}$. 
Therefore, if there are $n_{\rm clump}$ such clumps along the sight line 
where $E(J-H) \approx 0.23$ in Slit 4, we obtain a dust-to-gas ratio of 
$0.4-1 \times 10^{-3}~n_{\rm clump}^{-1}$.

The total flux of cool dust at 100 \micron\ in the global SED 
is $\sim 30 \pm 10$ Jy \citep{bar10}. 
The mass has been estimated to be $\sim 0.07~\msun$
if we adopt the FIR absorption coefficients for silicates from \citet{dor95}.
The contribution of small Fe or large Si grains to this mass budget
would depend on the volume filling factor of the clumps
but is probably not large. 
To address this issue, one may carry out an analysis
similar to that in this paper toward the entire ejecta shell.

%%%%%%%%%%%%%%%%%%%%%%%%%%%%%%%%%%%%%%%%%%%%%%%%%%%%%%%%%%%%%%%%%%%%%%%%%%%80==>
\section{Summary} \label{sec-sum}

We performed NIR spectroscopic observations of Cas A
in which we obtained the spectral and kinematical properties of 
63 \feii-line-emitting knots spread over the main ejecta shell. 
All of the knots show strong \feii\ 1.26 and 1.64 \micron\ lines,
the ratio of which provides a direct measure of the extinction.
From an analysis of the extinction toward individual knots,
we showed that the NIR extinction is due in part to SN dust within the remnant. 
We explored the nature of the SN dust responsible for the NIR extinction 
by analyzing its thermal infrared emission.  
Our main results are summarized below.

\begin{figure}
\center{
\includegraphics[scale=0.9]{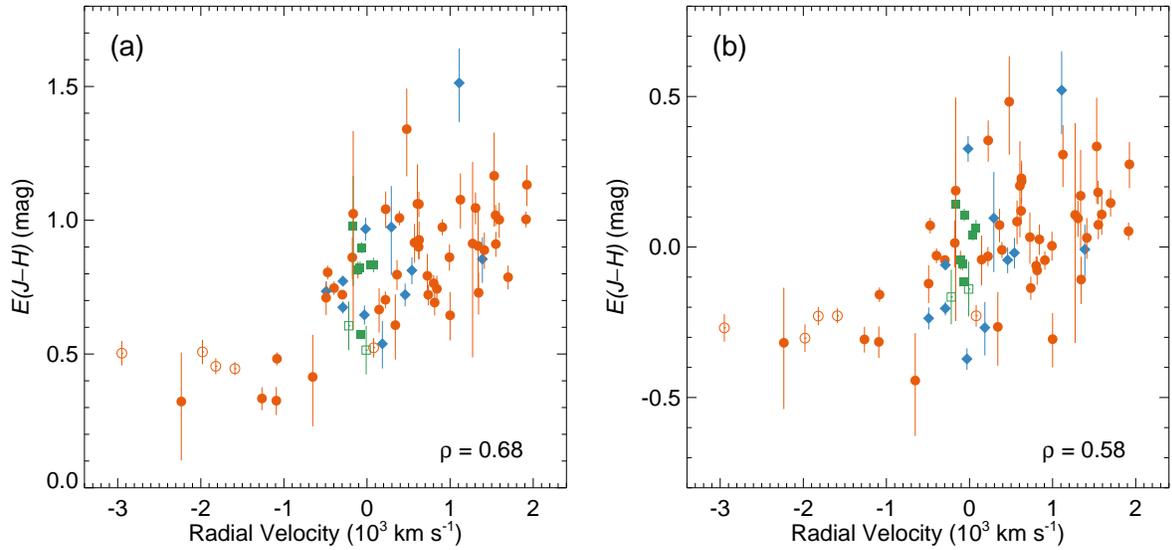}
}
\caption{
   (a) Variation of $E(J-H)$ for the 70 knots
   as a function of radial velocity. 
   Symbols and colors of the knots are the same as in Figure~\ref{fig-slit},
   and the correlation coefficient of the knots ($\rho$)
   is given in the bottom right.
   (b) Same as (a), but after the subtraction of
   the foreground extinction using the  $N_{\rm H}$ map of \citet{hwa12}.
} \label{fig-radvel}
\end{figure}

\noindent
1. We found that \feii\ emission from redshifted SN ejecta is 
in general more heavily obscured than that from blueshifted SN ejecta 
(Figure~\ref{fig-radvel}).
We interpret the correlation as evidence for
newly formed SN dust within the remnant. 
The amount of excess extinction varies considerably from one sight line to 
another, which suggests a highly non-uniform distribution of SN dust.

\begin{figure}
\center{
\includegraphics[scale=0.9]{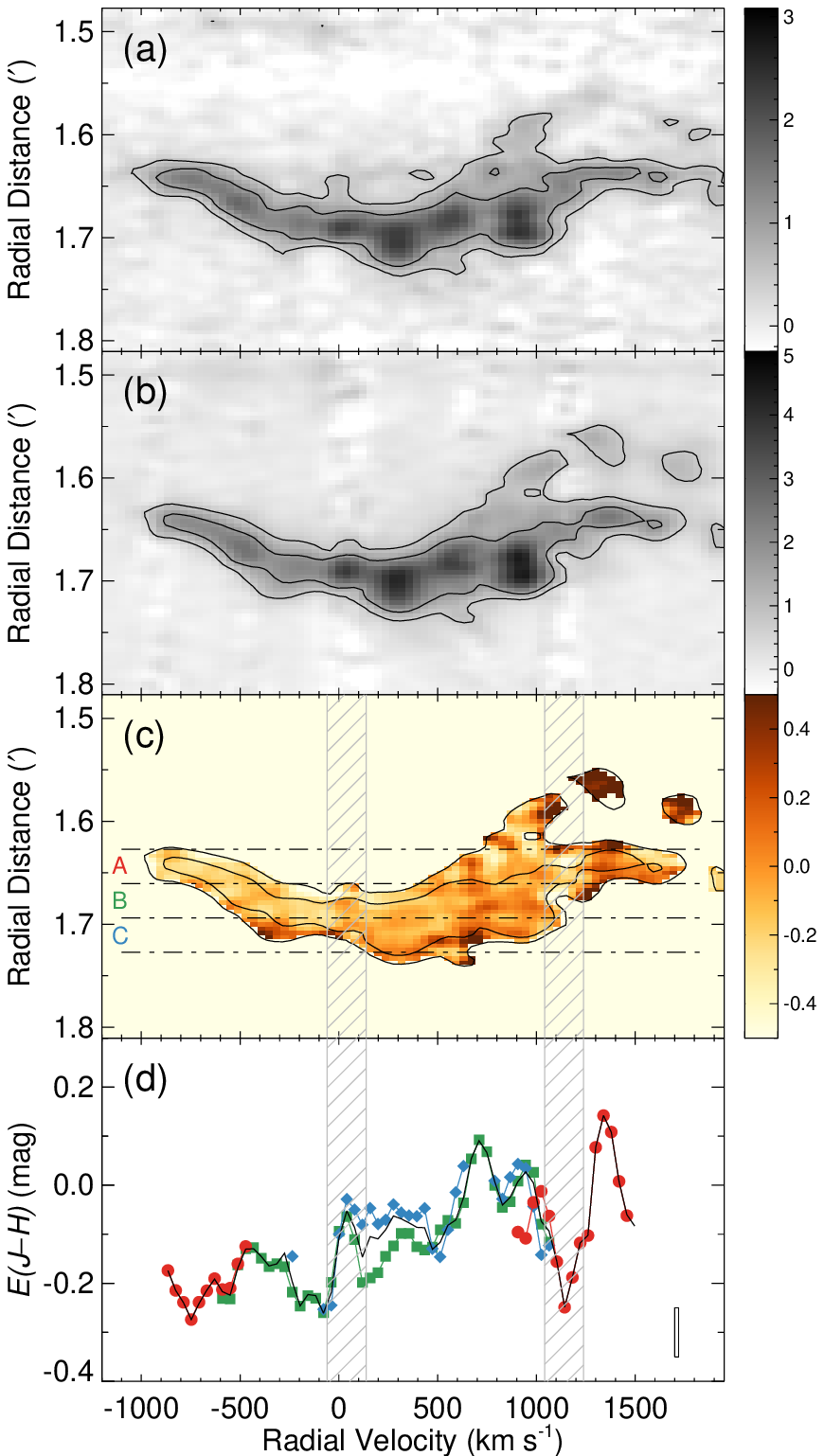}
}
\caption{
   Position-velocity diagram at
   (a) 1.26 \micron\ and (b) 1.64 \micron\ in Slit 4,
   showing the brightness of \feii\ 1.26 and 1.64 \micron\ emission.
   The y-axis represents the radial distance in arcminutes
   from the expansion center (see Figure~\ref{fig-slit}-(a)).
   The contour levels correspond to 2$\sigma$ and 4$\sigma$
   for \feii\ 1.26 \micron\
   and 3$\sigma$ and 6$\sigma$ for \feii\ 1.64 \micron\
   above the background rms noise.
   Unit of the color bar on the right side of the panels is
   $10^{-17}~{\rm erg~cm^{-2}~s^{-1}~\AA^{-1}}$.
   (c) Color excess ($E(J-H)$) map of Slit 4.
   The foreground extinction derived from the $N_{\rm H}$ map \citep{hwa12}
   is subtracted from the observed $E(J-H)$.
   The gray hatched areas are the areas
   where the OH line contamination is significant
   compared to the \feii\ emission.
   (d) One-dimensional $E(J-H)$ profile of the pixels above
   4$\sigma$ and 6$\sigma$ rms noise for \feii\ 1.26 and 1.64 \micron,
   respectively.
   The black solid line represents the intensity-weighted extinction profile,
   whereas the red, green, and blue solid lines are the mean profiles
   obtained from the slices in (c).
   The mean velocity resolution and extinction error are indicated by
   the black solid box at the lower right.
} \label{fig-pv5}
\end{figure}

\noindent
2. One slit (Slit 4) is located across the southern \feii\ ejecta filament,
which has the velocity structure of an expanding shell. 
Along this particular sight line, we measured an 
excess NIR extinction of $E(J-H)=0.23\pm0.05$ mag between the front and back sides  
of the shell, which must be entirely due to SN dust (Figure~\ref{fig-pv5}). 

\begin{figure}
\center{
\includegraphics[scale=0.9]{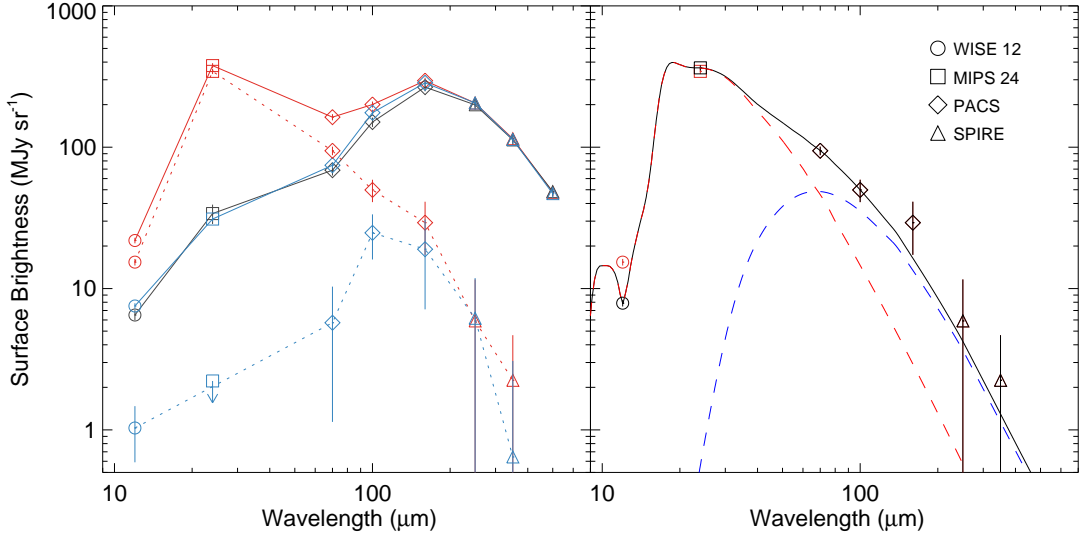}
}
\caption{
   (Left) Spectral energy distribution (SED) of dust emission in Slit 4. 
   The red and blue solid lines represent the SEDs of total dust continuum emission 
   in regions A and B in Figure~\ref{fig-profile},
   and the black solid line is that of the background region
   (see Table~\ref{tab-sed}).
   The red and blue dotted lines represent the background-subtracted SEDs in regions 
   A and B, respectively, and therefore the SEDs of pure SN dust.
   The symbol with a downward arrow denotes
   the 1$\sigma$ upper limit of the brightness.
   (Right) Two-component fit for the SED at position A. 
   The warm dust (red dashed line) is assumed to be MgSiO$_3$,
   whereas the cool dust (blue dashed line) is assumed to have 
   the general interstellar dust composition (see the text for details).
   Red and black symbols represent the measured and 
   color-corrected brightnesses, respectively.
} \label{fig-sed}
\end{figure}

\noindent
3. We analyzed the SED of thermal dust emission toward Slit 4 to show that 
there are warm ($\sim 100$ K) and cool ($\sim 40$ K) SN dust components
along the sight line (Figure~\ref{fig-sed}).
Among the grain species predicted from
theoretical dust formation models in CCSNe,
only silicate grains (i.e., MgSiO$_3$, Mg$_2$SiO$_4$, SiO$_2$)
can match the SED of warm SN dust which has a sharp 24 \micron\ peak.
The NIR extinction due to these warm silicate dust grains is negligible,
so the observed excess NIR extinction should be due to 
the cool SN dust.

\noindent
4. According to our analysis, 
only two grain species can explain both the SEDs of cool dust and 
the NIR excess extinction:
(1) small ($\lesssim 0.01~\micron$) Fe grains or
(2) large ($\gtrsim 0.1~\micron$) Si grains. 
However, neither would be consistent with the dust formation theory 
for SN IIb with uniformly expanding, layered ejecta \citep[e.g.,][]{noz10},
which predicts the formation of only 
small ($< 0.01~\micron$) Si grains and no Fe grains. 
However, if the SN ejecta is clumpy, we might expect that Fe grains and 
large Si grains can be produced in dense clumps.

\noindent
5. We suggest that the unshocked SN ejecta is clumpy and
the grain species in diffuse material and dense clumps are different, 
i.e., silicate grains in diffuse material
and small Fe or large Si grains in dense clumps.
In the shocked ejecta, both still exist, 
but it is the shock-heated warm grains in the diffuse ejecta
that emit most of the MIR and FIR ($\lesssim 70~\micron$) emission.
The cool dust that we detected could reside in either 
unshocked or shocked clumps,
and it is these cool grains that are responsible for the NIR extinction.
The contribution of each grain species to the total 
mass ($\gtrsim 0.1~\msun$) of cool SN dust in Cas A remains to be explored.

%%%%%%%%%%%%%%%%%%%%%%%%%%%%%%%%%%%%%%%%%%%%%%%%%%%%%%%%%%%%%%%%%%%%%%%%%%%80==>
\acknowledgments
We thank Takaya Nozawa for kindly providing his dust opacity curves and 
Una Hwang for providing the {\it Chandra} X-ray column density map.
We also thank Takashi Onaka​ and the referee Michael J. Barlow
for their helpful comments.
This research was supported by Basic Science Research Program through 
the National Research Foundation of Korea (NRF) funded by the Ministry of Science, 
ICT, and Future Planning (2014R1A2A2A01002811).
This research made use of Montage, funded by the National Aeronautics and Space
Administration's Earth Science Technology Office, Computation Technologies Project,
under Cooperative Agreement Number NCC5-626 between NASA and
the California Institute of Technology. Montage is maintained by
the NASA/IPAC Infrared Science Archive.

%%%%%%%%%%%%%%%%%%%%%%%%%%%%%%%%%%%%%%%%%%%%%%%%%%%%%%%%%%%%%%%%%%%%%%%%%%%80==>
{}

%%%%%%%%%%%%%%%%%%%%%%%%%%%%%%%%%%%%%%%%%%%%%%%%%%%%%%%%%%%%%%%%%%%%%%%%%%%80==>
\clearpage

\clearpage

\clearpage

\clearpage

\clearpage

\clearpage

\clearpage

\clearpage

\clearpage

%%%%%%%%%%%%%%%%%%%%%%%%%%%%%%%%%%%%%%%%%%%%%%%%%%%%%%%%%%%%%%%%%%%%%%%%%%%80==>
\clearpage
\begin{deluxetable}{lcccccc}
%\tabletypesize{\scriptsize}
\tabletypesize{\footnotesize}
%\rotate
%\setlength{\tabcolsep}{0.03in}
\tablewidth{0pt}
\tablecaption{\label{tab-sed}Observed and dust-continuum infrared brightnesses in Slit 4}
\tablehead{
\colhead{Waveband} &
\multicolumn{3}{c}{Observed\tablenotemark{a}}	&
\multicolumn{3}{c}{Dust-continuum}	\\
\colhead{}	&
\colhead{A} &
\colhead{B} &
\colhead{Background} &
\colhead{A} &
\colhead{B} &
\colhead{Background}
}
\startdata
WISE 12		&	24.0$\pm$1.1		&	7.9$\pm$0.4		&	6.6$\pm$0.5		&
			    21.9$\pm$1.0		&	7.5$\pm$0.3		&	6.5$\pm$0.5		\\
MIPS 24		&	397$\pm$16		&	37.1$\pm$1.5		&	34.6$\pm$5.3		&
			    378$\pm$15		&	31.0$\pm$1.2		&	33.9$\pm$5.3		\\
PACS 70		&	170$\pm$8		&	74.5$\pm$3.7		&	68.8$\pm$5.7		&
			    163$\pm$8		&	74.5$\pm$3.7		&	68.8$\pm$5.7		\\
PACS 100		&	210$\pm$11		&	175$\pm$9		&	151$\pm$11		&
			    201$\pm$10		&	175$\pm$9		&	151$\pm$11		\\
PACS 160		&	296$\pm$15		&	285$\pm$14		&	266$\pm$18		&
			    296$\pm$15		&	285$\pm$14		&	266$\pm$18		\\
SPIRE 250	&	205$\pm$14		&	206$\pm$14		&	200$\pm$15		&
			    205$\pm$14		&	206$\pm$14		&	200$\pm$15		\\
SPIRE 350	&	115$\pm$8		&	113$\pm$8		&	113$\pm$8		&
			    115$\pm$8		&	113$\pm$8		&	113$\pm$8		\\
SPIRE 500	&	48.0$\pm$3.4		&	47.0$\pm$3.3		&	48.5$\pm$3.4		&
			    48.0$\pm$3.4		&	47.0$\pm$3.3		&	48.5$\pm$3.4	
\enddata
\tablenotetext{a}{The observed brightnesses include contribution from emission lines.}
\tablecomments{
Brightnesses in units of MJy sr$^{-1}$.
The uncertainties are derived from a quadrature sum of the photometric uncertainty
and the brightness variation within the aperture.
}
\end{deluxetable}

\clearpage
\begin{deluxetable}{lccccc}
%\tabletypesize{\scriptsize}
\tabletypesize{\footnotesize}
%\rotate
%\setlength{\tabcolsep}{0.03in}
\tablewidth{0pt}
\tablecaption{\label{tab-grain}Dust grain species considered in this work}
\tablehead{
\colhead{Species} &
\colhead{Condition\tablenotemark{a}} &
\colhead{Type} &
\colhead{Density}  &
\colhead{$\lambda$ Coverage}   &
\colhead{References} \\
\colhead{} &
\colhead{} &
\colhead{} &
\colhead{(g cm$^{-3}$)}  &
\colhead{(\micron)}   &
\colhead{}
}
\startdata
C					&	u		&	Amorphous	&	2.28	&	4E-2 $-$ 2E3	&	1	\\
MgO					&	u		&	Crystalline	&	3.59	&	2E-3 $-$ 625	&	2	\\
MgSiO$_{3}$			&	m/u		&	Glassy		&	3.20	&	2E-1 $-$ 500	&	3	\\
Mg$_{2}$SiO$_{4}$	&	m/u		&	...			&	3.23	&	1E-1 $-$ 1E5	&	4	\\
Al$_{2}$O$_{3}$		&	m/u		&	Amorphous	&	4.01	&	2E-1 $-$ 500	&	5	\\
Si					&	u		&	Amorphous	&	2.34	&	1E-2 $-$ 148\tablenotemark{b}	&	6	\\
SiO$_{2}$			&	m/u		&	Amorphous	&	2.66	&	1E-4 $-$ 500	&	7	\\
Fe					&	u		&	...			&	7.95	&	1E-1 $-$ 1E5	&	4	\\
FeS					&	u		&	...			&	4.87	&	1E-1 $-$ 1E5	&	4	\\
Fe$_{3}$O$_{4}$		&	m		&	...			&	5.25	&	1E-1 $-$ 1E3	&	8
\enddata
\tablenotetext{a}{u = unmixed SN explosion model, m = mixed SN explosion model}
\tablenotetext{b}{
The maximum of $\lambda$ coverage is limited to 148 \micron.
In this case, we assume that the optical properties, $n$ and $k$, 
do not vary at longer wavelengths.}
\tablerefs{
(1) \citet{zub96},
(2) \citet{roe91},
(3) \citet{dor95},
(4) \citet{sem03},
(5) \citet{koi95} for $\lambda \le 8~\micron$ and \citet{beg97} for $\lambda > 8~\micron$,
(6) \citet{pil85},
(7) \citet{phi85},
(8) A. Triaud and H. Mutschke
(unpublished; see \url{http://www.astro.uni-jena.de/Laboratory/OCDB/mgfeoxides.html}).}
\end{deluxetable}

\clearpage
\begin{deluxetable}{cccccccc}
%\tabletypesize{\scriptsize}
\tabletypesize{\scriptsize}
%\rotate
%\setlength{\tabcolsep}{0.03in}
\tablewidth{0pt}
\tablecaption{\label{tab-param}Parameters of SN dust in Slit 4}
\tablehead{
\multicolumn{4}{c}{Warm Dust}	&
\multicolumn{4}{c}{Cool Dust}	\\
\hline
\colhead{Species} &
\colhead{$T$} &
\colhead{$\Sigma_{\rm d}$} &
\colhead{$E(J-H)$}  &
\colhead{Species}   &
\colhead{$T$} &
\colhead{$\Sigma_{\rm d}$} &
\colhead{$E(J-H)$} \\
\colhead{} &
\colhead{(K)} &
\colhead{(10$^{-6}$ g cm$^{-2}$)} &
\colhead{(10$^{-4}$ mag)}  &
\colhead{}   &
\colhead{(K)} &
\colhead{(10$^{-4}$ g cm$^{-2}$)} &
\colhead{(mag)}
}
\startdata
&	\multirow{4}{*}{$115\pm2$}	&	
	\multirow{4}{*}{$0.032\pm0.005$}	&	\multirow{4}{*}{$0.11\pm0.02$}	&	
	Si	&	\multirow{2}{*}{$45\pm6$}	&	
	\multirow{2}{*}{$0.039\pm0.022$}	&	\multirow{2}{*}{$0.15-0.65$}	\\
MgSiO$_3$	&	&	&	&	
	($0.16 \leq a \leq 0.21\micron$)	&	&	&	\\
($a=0.1\micron$)	&	&	&	&	
	Fe	&	\multirow{2}{*}{$\leq45\pm7$}	&	
	\multirow{2}{*}{$\geq2.0\pm1.3$}	&	\multirow{2}{*}{$\geq0.16\pm0.10$}	\\
&	&	&	&
	($a\leq0.01\micron$)	&	&	&	\\
\hline
&	\multirow{4}{*}{$92\pm1$}	&	
	\multirow{4}{*}{$0.14\pm0.02$}	&	\multirow{4}{*}{$0.60\pm0.09$}	&	
	Si	&	\multirow{2}{*}{$30\pm7$}	&	
	\multirow{2}{*}{$0.054\pm0.052$}	&	\multirow{2}{*}{$0.21-0.90$}	\\
Mg$_2$SiO$_4$	&	&	&	&	
	($0.16 \leq a \leq 0.21\micron$)	&	&	&	\\
($a=0.1\micron$)	&	&	&	&	
	Fe	&	\multirow{2}{*}{$\leq26\pm6$}	&	
	\multirow{2}{*}{$\geq6.9\pm7.1$}	&	\multirow{2}{*}{$\geq0.55\pm0.56$}	\\
&	&	&	&
($a\leq0.01\micron$)	&	&	&	\\
\hline
&	\multirow{4}{*}{$78\pm1$}	&	
	\multirow{4}{*}{$0.9\pm0.1$}	&	\multirow{4}{*}{$2.2\pm0.3$}	&
	Si	&	\multirow{2}{*}{$42\pm7$}	&	
	\multirow{2}{*}{$0.035\pm0.025$}	&	\multirow{2}{*}{$0.14-0.59$}	\\
SiO$_2$	&	&	&	&
	($0.16 \leq a \leq 0.21\micron$)	&	&	&	\\
($a=0.1\micron$)	&	&	&	&	
Fe	&	\multirow{2}{*}{$\leq40\pm7$}	&	
	\multirow{2}{*}{$\geq2.1\pm1.8$}	&	\multirow{2}{*}{$\geq0.17\pm0.14$}	\\
&	&	&	&
	($a\leq0.01\micron$)	&	&	&	
\enddata
\end{deluxetable}

\end{document}